\documentstyle[12pt]{article}

\def\[{\left\lbrack}
\def\]{\right\rbrack}

\def\({\left(}
\def\){\right)}
\def\ih{\'\i}

\title{Does the Weyl ordering prescription lead to the correct
energy levels for the quantum particle on the
D-dimensional sphere ?}

\author{Jorge Ananias Neto\thanks{e-mail:jorge@fisica.ufjf.br}
 and Wilson Oliveira\thanks{e-mail:wilson@fisica.ufjf.br}\\
Departamento de F\ih sica, ICE \\ Universidade Federal de Juiz de
Fora, 36036-330, \\ Juiz de Fora, MG, Brazil }

\date{}

\begin{document}

\maketitle

\begin{abstract}
\noindent The energy eigenvalues of the quantum particle constrained in
a surface of the sphere of D dimensions embedded in a $R^{D+1}$ space are
obtained by using two different procedures: in the first, we derive the 
Hamiltonian operator by squaring the expression of the momentum, 
written in cartesian components,
which satisfies the Dirac brackets between the canonical operators of 
this second class  system.
We use the Weyl ordering prescription to construct the Hermitian
operators. When $D=2$ we verify that there is no constant
parameter in the expression of the eigenvalues energy, a result that is 
in agreement with the fact that an extra term would change the level
spacings in the hydrogen atom; in the second procedure it is  
adopted the non-abelian BFFT
formalism to convert the second class constraints into first class ones.
The non-abelian first class Hamiltonian operator is symmetrized by also 
using the Weyl ordering rule. We observe that their energy eigenvalues
differ from a constant parameter when we compare with the second class
system. Thus, a conversion of the D-dimensional sphere second class 
system for a first class one does not reproduce the same values.

\end{abstract}

\maketitle

\setlength{\baselineskip} {20 pt}

\section{Introduction}
When one obtains the representation of the physical operators
given by the algebraic expressions of the Dirac commutators 
\footnote{\noindent These commutators define the quantum structure 
of a particular theory.}of a specific theory we must pay attention if 
these canonical operators are Hermitian ones. If it is not true, this
fact indicates that there are the so-called operator ordering 
problems. It is well known that many theories, for example, we
can cite the non-linear sigma model\cite{Banerjee3},
the Skyrme model\cite{Skyrme}, D-dimensional sphere quantum mechanics
\cite{Abdalla, Kleinert}, which will be the subject of this paper, 
present ordering problems. 
The ordering problems, in many cases, appear as a consequence of the 
constraints dynamic acting in the systems. Therefore, the presence of the 
constraints can lead to a non-trivial relations for the Dirac commutators 
which lead to an operator 
level problems for the representations of the physical operators of a 
peculiar theory. 

Some articles are devoted to study the problem of the quantization of 
the free particle constrained in a surface of D-dimensional sphere. 
We must mention the works of 
Abdalla and Banerjee\cite{Abdalla}
who used a Lagrangian reduced method to obtain the quantum Hamiltonian
of the multidimensional rotor;
Kleinert and Shabanov\cite{Kleinert} who 
employed the angular momentum algebra operators to calculate the energy
spectrum; Foerster, Girotti and Kuhn\cite{Foerster} who used the metrical formalism to described the quantum 
particle in a curved space.

The purpose of this paper is to calculate with more rigour the energy
spectrum of a quantum particle lying in a surface of D-dimensional
sphere. We use the second and first class Dirac methods of
quantization\cite{Dirac}. In both cases, the Weyl prescription is 
employed to construct the physical Hermitian operators. 

This paper is organized as follows. In Section 2, we derive the 
energy levels of the quantum Hamiltonian constructed from the Hermitian momentum operator, written in a rectangular coordinates, which 
satisfies the Dirac brackets. For D=2  we verify that this energy levels 
are in accordance with experimental values\cite{Kleinert}. This 
result can suggest the
power of the Dirac method quantization of the constrained systems. 
In Section 3, we convert the 
second-class constraints into the first class ones. For this, we use the 
non-abelian BFFT formalism\cite{Banerjee1},\cite{nabft}. 
With the first class Hamiltonian we calculate the
energy eigenvalues of the quantum particle which differ by a constant 
parameter from the obtained by Dirac brackets. In Section 4, we give the 
conclusions. In the Appendices we perform a brief review about the 
non-abelian BFFT procedure, where we calculate the Lagrangian of this 
new first class system.

\section{Dirac brackets for the free particle on the D-dimensional
 sphere}
The quantization of the second class constrained systems 
is usually performed by using the method proposed by Dirac, Bergman
and co-workers\cite{Dirac}. 
The constraints are classified as primary and secondary ones. 
Secondary constraints are obtained from
the condition that primary constraints are conserved in time.
One must  repeat the condition requiring time derivative
of secondary constraints vanish until all independent constrains
are obtained. If the whole second class constraints are established, 
the so called Dirac bracket for the canonical variables A and B 
is given by

\begin{equation}
\label{formula 1}
\{A,B \}^*= \{A,B \}-\{A,\phi_\alpha \}C^{-1}_{\alpha\beta}
\{\phi_\beta,B \},
\end{equation}

\noindent where $\phi_\alpha$ and $\phi_\beta$ are the second class
constraints and the matrix elements $ C_{\alpha\beta}$ is defined by

\begin{equation}
\label{formula 2}
C_{\alpha \beta}=\{\phi_\alpha,\phi_\beta \}.
\end{equation}

\noindent The quantum mechanics commutators are given by the
replacement \break $\{\,,\,\}_D \rightarrow -i\[\,,\,\]$.

Now, let us consider the dynamic of a particle in the D-sphere 
manifolds. The primary constraint is

\begin{equation}
\label{formula 3}
\phi_1=x_ix_i-R^2=0,
\end{equation}

\noindent where R is the radius of the sphere.
With the expression of the classical Hamiltonian given by

\begin{equation}
\label{formula 4}
H_c={1\over 2} p_i p_i \,\,,
\end{equation}

\noindent we obtain the secondary constraint

\begin{equation}
\label{formula 5}
\phi_2=x_ip_i=0,
\end{equation}

\noindent which expresses the fact that a motion on the surface of sphere 
has no radial component.
From the expressions of primary and secondary constraints,
(\ref{formula 3}) and (\ref{formula 5}), we obtain the algebra
of the canonical operators
which defines the quantum mechanics of a particle in a surface of
D-dimensional sphere\cite{Kleinert},\cite{JAN}

\begin{eqnarray}
\label{formula 6}
\[ x_i,x_j \]& = & 0,\\
\label{formula 7}
\[ x_j,p_k \]& = & i \( \delta_{jk} - {x_j x_k \over R^2} \),\\
\label{formula 8}
\[ p_j,p_k \]& = &  {i\over R^2}(p_j x_k-p_k x_j).
\end{eqnarray}

\noindent Here we would like to comment the operator ordering
problem occurring in the right-hand side of Eq.~(\ref{formula 8}).
This fact is solved under condition that this commutator
satisfied the equation\footnote{We would like to acknowledge
M. Plyushchay for the remarks about this question.}  
$\, \[ p_j,p_j \] = 0 $ .

The solution of the momentum operator that satisfies the 
canonical relations (\ref{formula 7}) and (\ref{formula 8}) is

\begin{equation}
\label{formula 9}
p_k={1\over i} \[ \partial_k - {x_k x_i \over R^2}\partial_i \].
\end{equation}

\noindent The Laplacian operator in rectangular coordinates is
\footnote{ The Laplacian expression in $D+1$ dimension written in 
spherical coordinates is given by\cite{Jaber}\\
$\nabla^2 = {1\over R^2}\, 
( \,\,\sum_{K=1}^{D-1}( \prod_{j=1}^K \sin \theta_j)^{-2}
(\sin \theta_K)^{K+2-D} {\partial\over \partial \theta_K}
(\sin \theta_K^{D-K} {\partial\over \partial \theta_K} )\\
+ ( \prod_{j=1}^{D-1} \sin \theta_j )^{-2}
{\partial^2 \over \partial \varphi^2} \,\, )$. 
}

\begin{equation}
\label{rectangular}
p_k\cdot p_k = \partial_k\partial_k 
+ {1\over R^2}\( Op Op + (D-1)Op \) ,
\end{equation}

\noindent where the operator $Op$ is defined as $Op\equiv x_i\partial_i$
and D is the dimension of the sphere embedded in a $D+1$ 
dimensional cartesian space\footnote{\noindent Notice that 
$\partial_i x_i= \delta_{ii}=D+1$.}.
The eigenstates are symmetric polynomials
in the $x_i's$~\cite{Adkins} defined as 
$|polynomial\rangle ={1\over N(l)}(x_m+ i x_n)^l\,$
, where $N(l)$ is a
normalization factor, $m$ and $n$ take values $m,n=1,2,\cdots D+1 $, 
and $l$ is an integer parameter\footnote{
\noindent The eigenvalues of the operator Op are defined by the following
equation:\\ $Op |polynomial\rangle = l |polynomial\rangle$.}, 
$l=0,1,2, ...\,\, $.
The expression $p_k \cdot p_k$ is known as the laplacian
$\nabla^2$ on the D-sphere.

We must mention the problem of ordering that
appear in Eq.~(\ref{formula 9}). Due to this fact, the canonical
momentum $p_k$ is not a Hermitian operator as required by quantum 
mechanics. However, we can solve this problem by using the 
Weyl ordering operator prescription\cite{Weyl}. In the case of the 
canonical momentum, Eq.~(\ref{formula 9}), this rule expresses 
that the new operator must be constructed by counting all possible 
randomly order of the $x's$ and $\partial\,\,$. Then, the symmetric
momentum operator $p_k$ reads

\begin{eqnarray}
\label {formula 10}
[p_k]_{sym} & = &{1\over 6i} (6\partial_k
-{1\over R^2}x_k x_i\partial_i- {1\over R^2}x_k\partial_i x_i
-{1\over R^2}x_i x_k\partial_i-{1\over R^2}x_i\partial_i x_k\nonumber\\ 
&-&{1\over R^2}\partial_i x_k x_i
-{1\over R^2}\partial_i x_i x_k)\nonumber\\ \nonumber \\
& = & {1\over i} \( \partial_k-{1\over R^2}x_k x_i \partial_i
-\( {D+2\over 2R^2} \)x_k \).
\end{eqnarray}

\noindent The quantum Hamiltonian is obtained performing the square of the 
Hermitian operator $[p_k]_{sym}$

\begin{eqnarray}
\label{formula 11}
H&=&{1\over 2}~[p_k]_{sym}\cdot[p_k]_{sym}\nonumber\\
&=&-{1\over 2}\partial_k\partial_k
+{1\over 2R^2}\( Op Op + (D-1)Op +{D^2-4\over 4} \), 
\end{eqnarray}  

\noindent where the operator $Op$ is defined in ref.(\ref{rectangular}).
Thus, applying the
Hamiltonian operator on the physical states $|polynomial\rangle$ we
obtain the energy levels

\begin{eqnarray}
\label{formula 12}
E_l={1\over 2R^2} \[ l (l+D-1) + {D^2-4\over 4} \].
\end{eqnarray}  

\noindent We would like to point out that the energy formula
(\ref{formula 12}) gives no additional constant energy for
a particle on a circle\footnote{For D=3 the extra term in
Equation~(\ref{formula 12}) is the same obtained in the three-sphere 
collective coordinates Skyrmions quantization\cite{JAN2}.}, D=2.
This result is in agreement with the experimental energy level spacings
of the hydrogen atom. Thus, for D=2, the Dirac quantization procedure 
together with the Weyl ordering prescription  predict the correct energy 
levels for the quantum particle lying on the sphere.

\section {The non-abelian BFFT formalism for the 
free particle on the D-dimensional sphere}

In the appendix A we perform a brief review of the BFFT formalism
and its non-abelian extension. In the appendix B we obtain the
Lagrangian of this new theory.

\vskip .5 cm

In the Section 2, we have seen the Hamiltonian 
for a particle of unit mass moving on the surface of a D-dimensional 
sphere of  radius R isometrically embedded in $R^{D+1}$ is 
given by

\begin{equation}
\label{formula 41}
H_c={1\over 2} p_i p_i \,\,,
\end{equation}

\noindent with the primary and the secondary constraints respectively
written as

\begin{eqnarray}
\label{formula 31}
\phi_1=x_ix_i-R^2=0,\\
\phi_2=x_ip_i=0.
\end{eqnarray}

\noindent The constraints $T_1$ and $T_2$ are of the second class. 
The matrix elements of their Poisson brackets read

\begin{equation}
\label{3.5}
\Delta_{\alpha \beta} = \{T_\alpha,T_\beta\} = -2 \epsilon_{\alpha \beta}
x_ix_i \, , \,\, \alpha,\beta = 1,2
\end{equation}

\noindent where $\epsilon_{\alpha \beta}$ is the antisymmetric
tensor normalized as $\epsilon_{12} = -\epsilon^{12} = -1$.

To implement the extended non-abelian BFFT formalism, we introduce 
auxiliary coordinates, one for each of the second class constraint.
Let us generally denote them by $\eta^\alpha$, where $\alpha=1,2$, 
and consider that the Poisson algebra of these new coordinates 
is given by

\begin{equation}
\label{3.6}
\{ \eta^\alpha, \eta^\beta \} = \omega^{\alpha \beta}
= 2\epsilon^{\alpha\beta}; 
\,\,\alpha=1,2.
\end{equation}

\noindent From Eq.~(\ref{A26}), we have

\begin{equation}
\label{3.7}
2X_{11}\,X_{22} = -2\,x_i x_i\ + \,C_{12}^1\,T_1.
\end{equation}

\noindent
After some attempts, we find that a convenient choice for these 
coefficients is

\begin{eqnarray}
&&X_{11}=R,
\nonumber\\
&&X_{22}=-R,
\nonumber\\
&&X_{12}=0=X_{21},
\nonumber\\
&&C_{12}^1=2,
\nonumber\\
&&C_{12}^2=0.
\label{3.8}
\end{eqnarray}

\noindent Using (\ref{A4}), (\ref{A6}), (\ref{A11}), (\ref{3.6}) and
(\ref{3.8}), the new set of constraints is found to be

\begin{eqnarray}
\label{3.9}
\tilde{T}_1=x_i x_i-R^2+R\eta^1,\\
\label{3.10}
\tilde{T}_2=x_ip_i-R\eta^2+\eta^1\eta^2.
\end{eqnarray}

\noindent The first class constraint algebra is

\begin{eqnarray}
&&\bigl\{\tilde T_1,\,\tilde T_1\bigr\}=0,
\nonumber\\
&&\bigl\{\tilde T_1,\,\tilde T_2\bigr\}
=2\,\tilde T_1,
\nonumber\\
&&\bigl\{\tilde T_2,\,\tilde T_2\bigr\}=0.
\label{3.11}
\end{eqnarray}

Next, we derive the corresponding Hamiltonian in the extended
phase space. The corrections for the canonical Hamiltonian are
given by Eqs. (\ref{A20}) and (\ref{A28}). With the objective
to simplify the expression of the first class Hamiltonian, we
chose an algebra for the system defined by the parameters
$B_a^b$ in (\ref{A27}). We have verified that possible values are

\begin{equation}
\label{3.12}
B_1^1 = B_1^2=B_2^1=B_2^2=0.
\end{equation}

\noindent Using the inverse matrices

\begin{eqnarray}
\label{3.13}
\omega_{\alpha\beta} = {1\over 2} \epsilon_{\alpha\beta}, \\\
\label{3.14}
X^{\alpha \beta} = \left( \begin{array}{clcr} 1\over R  & \,\,0 \\ 0
& -{1\over R}\end{array} \right),
\end{eqnarray}

\noindent and the algebra defined by (\ref{3.12}), it is possible
to compute the involutive first class Hamiltonian

\begin{eqnarray}
\label{3.15}
\tilde{H}={1\over 2} p_ip_i(1-{\eta^1\over R})
-x_ip_i{\eta^2\over R}(1-{\eta^1\over R})
+ {1\over 2} x_i x_i{\eta^2\eta^2\over R^2}(1-{\eta^1\over R}),
\end{eqnarray}

\noindent which satisfies the first class Poisson algebra

\begin{eqnarray}
\label{3.16}
&\{\tilde{T}_1, \tilde{H}_2\} = 0,\,\,
\,\,\,\,(B_1^1=B_1^2=0)\\
\label{3.17}
&\{\tilde{T}_2, \tilde{H}_2 \} = 0. \,\,\,\,\,\,\,\,\,\,\,\,\,\,
\,\,\,\,\,\,\,\,
(B_2^1=B_2^2=0)
\end{eqnarray}

\noindent Here we would like to remark that, contrary the results
obtained by the abelian BFFT method applied to the non-linear
Lagrangian theories \cite{Banerjee3,WOJAN}, the expression of the
first class Hamiltonian (\ref{3.15}) is a finite sum. As it was 
emphasized in the introduction, the possibility 
pointed out by Banerjee, Banerjee and Ghosh to obtain non-abelian
first class theories leads to a  more elegant and  simplified 
Hamiltonian structure than usual abelian BFFT case.

    Here we intend to obtain the spectrum of the extended theory.
We use the Dirac method of quantization for the first class constraints
\cite{Dirac}.The basic idea consists in imposing quantum 
mechanically the first class constraints as operator 
condition on the wave-functions as a way to obtain the physical 
subspace, i.e.,

\begin{equation}
\label{qope}
\tilde{T}_\alpha | \psi \rangle_{phys} = 0, \,\,\,\, \alpha=1,2.
\end{equation}

\noindent The operators $\tilde{T}_1\,$ and $\tilde{T}_2\,$
are

\begin{eqnarray}
\label{qope1}
\tilde{T}_1=x_ix_i-R^2+R\eta^1,\\
\label{qope2}
\tilde{T}_2=x_ip_i - R\eta^2+\eta^1\eta^2.
\end{eqnarray}

\noindent Thus, the physical states that satisfy (\ref{qope}) are

\begin{equation}
\label{physical}
| \psi \rangle_{phys} = {1\over V } \, \delta (x_ip_i
- R\eta^2+\eta^1\eta^2) \,\delta(x_i x_i-R^2+R\eta^1)\,|polynomial \rangle,
\end{equation}

\noindent where {\it V } is the normalization factor 
and the ket {\it polynomial} is defined by
$|polynomial \rangle ={1\over N(l)} (x_1+ i x_2)^l \,$. The 
corresponding quantum Hamiltonian of (\ref{3.15}) )
will be indicated as 

\begin{eqnarray}
\label{echs1}
\tilde{H}={1\over 2} p_ip_i(1-{\eta^1\over R})
-x_ip_i{\eta^2\over R}(1-{\eta^1\over R})
+ {1\over 2} x_i x_i{{\eta^2\eta^2}\over R^2}(1-{\eta^1\over R}).
\end{eqnarray}

\noindent Thus, in order to obtain the spectrum of the theory, we take
 the scalar product, 
$_{phys}\langle\psi| \tilde{H} | \psi \rangle_{phys}\,$,
that is the mean value of the extended Hamiltonian. Then

\begin{eqnarray}
\label{mes1}
_{phys}\langle\psi| \tilde{H} | \psi \rangle_{phys}=\nonumber \\
\langle polynomial |\,\,  {1\over V^2} \int d\eta^1 d\eta^2 
\delta(x_i x_i - R^2 + R\eta^1)\delta(x_ip_i - R\eta^2+\eta^1
\eta^2)\nonumber \\
\tilde{H}
\delta(x_ip_i - R\eta^2+\eta^1\eta^2)\delta(x_i x_i - R^2 + R\eta^1)\,\,
| polynomial \rangle .
\end{eqnarray}

\noindent Notice that due to the presence of the delta functions
 $\delta(x_i x_i - R^2 + R\eta^1)$ and
$\delta(x_ip_i - R\eta^2+\eta^1\eta^2)$ in
(\ref{mes1}) the scalar product can be simplified. 
Then, integrating over $\eta^1$ and $\eta^2$ we obtain\footnote{The 
regularization of delta function squared like
$(\delta(x_i x_i - R^2 + R\eta^1))^2$ and 
$(\delta(x_i p_i - R\eta^2+\eta^1\eta^2))^2$ 
is performed by using the delta relation, $(2\pi)^2\delta(0)=
\lim_{k\rightarrow 0}\int d^2x \,e^{ik\cdot x} =\int d^2x= V.$ 
Then, we use the parameter V as the normalization factor.}

\begin{eqnarray}
\label{mes13}
_{phys}\langle\psi| \tilde{H} | \psi \rangle_{phys}=\nonumber \\
\langle polynomial |  {1\over 2R^2}  x_ix_i p_j p_j 
- {1\over 2R^2} x_i p_i x_j p_j | polynomial \rangle .
\end{eqnarray}

\noindent The final Hamiltonian operator inside the kets (\ref{mes13})
 must be Hermitian. Then, this Hamiltonian has to be 
symmetrized. Following the prescription of Weyl ordering\cite{Weyl}
(symmetrization procedure) we can write the symmetric 
Hamiltonian as

\begin{eqnarray}
\label{HWeyl}
\tilde{H}_{sym} = {1\over 2R^2}  \[ x_ix_i p_j p_j \]_{sym} 
- {1\over 2R^2}\[ x_ip_i x_j p_j \]_{sym},
\end{eqnarray}

\noindent where $\[ x_ix_i p_j p_j \]_{sym} $ and
 $\[ x_ip_i x_jp_j \]_{sym}$ are defined as

\begin{eqnarray}
\label{wdef}
\[ x_i x_i p_j p_j \]_{sym} = {1\over 24}
( 4x_ix_ip_jp_j+4x_ip_jx_ip_j+4x_ip_jp_jx_i \\
+4p_jx_ix_ip_j+4p_jx_ip_jx_i+4p_jp_jx_ix_i ), \\
\[ x_ip_i x_jp_j \]_{sym} = {1\over 24} 
(2x_ip_ix_jp_j+2x_ip_ip_jx_j+2x_ip_jp_ix_j \\
+2x_ip_jx_jp_i+2x_ix_jp_jp_i+2x_ix_jp_ip_j \\
+2p_ix_ix_jp_j+2p_ix_ip_jx_j+2p_ip_jx_ix_j \\
+2p_ip_jx_jx_i+2p_ix_jp_jx_i+2p_ix_jx_ip_j ).
\end{eqnarray}

\noindent Then, using the symmetric Hamiltonian operator 
$\tilde{H}_{sym}$,  eq.~(\ref{HWeyl}), the mean value 
(\ref{mes13}) is

\newpage

\begin{eqnarray}
\label{mes2W}
_{phys}\langle\psi| \tilde{H}_{sym} | \psi \rangle_{phys}=\nonumber \\
\langle polynomial | {1\over 2R^2}  \[ x_ix_i p_j p_j \]_{sym} 
- {1\over 2R^2}\[ x_ip_i x_jp_j \]_{sym}| polynomial \rangle .
\nonumber\\
=\langle polynomial | {1\over 48R^2}
( 4x_ix_ip_jp_j+4x_ip_jx_ip_j+4x_ip_jp_jx_i+4p_jx_ix_ip_j \\
+4p_jx_ip_jx_i+4p_jp_jx_ix_i )\nonumber \\- {1\over 48R^2} 
(2x_ip_ix_jp_j+2x_ip_ip_jx_j+2x_ip_jp_ix_j+2x_ip_jx_jp_i+2x_ix_jp_jp_i \\
+2x_ix_jp_ip_j+2p_ix_ix_jp_j+2p_ix_ip_jx_j+2p_ip_jx_ix_j+2p_ip_jx_jx_i \\
+2p_ix_jp_jx_i+2p_ix_jx_ip_j )
| polynomial \rangle \,\,.
\end{eqnarray}

\noindent The operator $\pi^j$ describes the momentum of free particle 
and its representation on the collective coordinates space $x_i$ 
is given by

\begin{equation}
\label{piconfig}
\pi^j = -i {\partial\over \partial x_j}\,\,.
\end{equation}

\noindent Substituting the expression (\ref{piconfig}) into 
(\ref{mes2W}), we obtain the energy levels, read as

\begin{eqnarray}
\label{meswe}
E_l = _{phys}\langle\psi| [\tilde{H}]_{sym} | \psi \rangle_{phys}  = 
 {1\over 2R^2} \[ l(l+D-1) +  {D(D+1)\over 4} \] \,\,.
\end{eqnarray}

\noindent Comparing with the second-class energy expression, 
formula (\ref{formula 12}), we see that this expression differs by a 
constant parameter\footnote{Recently, a work of Hong, Kim and Park\cite{Hong}
suggest a generalized momenta definition that can conciliate the energy spectra obtained by Dirac and BFT formalisms, in the context of collective
coordinates quantization of the Skyrme model.}

\section{Conclusions}

In this work we perform a study about the energy spectrum
of a quantum particle constrained in a surface of the D-dimensional
sphere. We use two procedures to obtain the D-sphere energy eigenvalues: in the first, we calculate the Hamiltonian by 
squaring the Hermitian momentum operator. This symmetrical momentum 
is derived, firstly, by an explicit solution of the Dirac brackets between the canonical operators, and then, employing the Weyl ordering prescription, we obtain the Hermitian momentum operator. It is clear 
that the square of this operator (that is proportional to the Hamiltonian) 
also is Hermitian one. When we put D=2 (the physical system) 
the energy eigenvalues are according to experimental results. The second procedure is the conversion of the second class constraints of the D-sphere system into first class ones. We use an extension of the BFFT formalism. 
After this, we also employ the Weyl ordering rules to obtain the first class Hermitian Hamiltonian. Its eigenvalues differs by a constant parameter compared with the spectrum energy in the second class method.
Thus, in the case of D-dimensional sphere quantum mechanics, Dirac Brackets
quantization of the second class constraints can lead to the correct 
experimental results, contrary to the predict by the non-abelian BFFT 
formalism.

\section{Acknowledgments}
We would like to thank F.I.Takakura for critical reading.This work is supported in part by FAPEMIG, Brazilian Research Council.

\section{Appendix A: Brief review of the BFFT formalism and its non-abelian 
extension}

\renewcommand{\theequation}{A\arabic{equation}}
\setcounter{equation}{0}

The BFFT formalism of converting the second class constraints
into the first class ones is a recent procedure\footnote{
Together with the Dirac first class quantization 
prescription.} used to quantize systems
of second class constraints\footnote{\noindent Compared with the 
well known Dirac brackets method.}. Then, the purpose of this appendix
is to exhibit a review of the BFFT formalism.

Let us consider a system described by a Hamiltonian $H_0$ in a
phase space $(q^i,p^i)$ with $i=1,\dots,N$. It is supposed that there the system possesses only the second class constraints. Denoting them by $T_a$, with $a=1,\dots ,M<2N$,
we arrive at the following algebra

\begin{equation}
\bigl\{T_a,\,T_b\bigr\}=\Delta_{ab},
\label{A1}
\end{equation}

\noindent
where $\det(\Delta_{ab})\not=0$.

As it was mentioned above, the general purpose of the BFFT formalism
is to convert the second class constraints into the first class ones. 
This goal is achieved by
introducing canonical variables, one for each of the second class constraint
(the connection between the number of the second class constraints and
the number of the new variables should be equal in order to preserve
the same number of the physical degrees of freedom in the resulting extended
theory). We denote these auxiliary variables by $\eta^a$ and assume
that they satisfy the following algebra

\begin{equation}
\bigl\{\eta^a,\,\eta^b\bigr\}=\omega^{ab}.
\label{A2}
\end{equation}

\noindent
Here $\omega^{ab}$ is a constant non-degenerate matrix
(~det$(\omega^{ab})\neq 0$~).
The obtainment of $\omega^{ab}$ is embodied in the calculation 
of the resulting first class constraints which are denoted as 
$\tilde T_a$. Of course, these constraints depend on the new
variables $\eta^a$, that is

\begin{equation}
\tilde T_a=\tilde T_a(q,p;\eta),
\label{A3}
\end{equation}

\noindent 
and are supposed to satisfy the boundary condition

\begin{equation}
\tilde T_a(q,p;0)= T_a(q,p).
\label{A4}
\end{equation}

\noindent
In the framework of the BFFT formalism, the characteristic property of
the new constraints is that they are assumed to be strongly
involutive, i.e.

\begin{equation}
\bigl\{\tilde T_a,\,\tilde T_b\bigr\}=0.
\label{A5}
\end{equation}

\noindent
The solution of Eq.~(\ref{A5}) can be achieved by considering
$\tilde T_a$ expanded as

\begin{equation}
\tilde T_a=\sum_{n=0}^\infty T_a^{(n)},
\label{A6}
\end{equation}

\noindent
where $T_a^{(n)}$ is a term of order $n$ in $\eta$. The condition of compatibility with the boundary condition~(\ref{A4}) requires 

\begin{equation}
T_a^{(0)}=T_a.
\label{A7}
\end{equation}

\noindent
Substituting the Eq.(\ref{A6}) into (\ref{A5}) leads to a set of
equations, one for each coefficient of $\eta^n$. We list some of them
below

\newpage

\begin{eqnarray}
&&\bigl\{T_a,T_b\bigr\}
+\bigl\{T_a^{(1)},T_b^{(1)}\bigr\}_{(\eta)}=0
\label{A8}\\
&&\bigl\{T_a,T_b^{(1)}\bigr\}+\bigl\{T_a^{(1)},T_b\bigr\}
+\bigl\{T_a^{(1)},T_b^{(2)}\bigr\}_{(\eta)}
+\bigl\{T_a^{(2)},T_b^{(1)}\bigr\}_{(\eta)}=0
\label{A9}\\
&&\bigl\{T_a,T_b^{(2)}\bigr\}
+\bigl\{T_a^{(1)},T_b^{(1)}\bigr\}_{(q,p)}
+\bigl\{T_a^{(2)},T_b\bigr\}
+\bigl\{T_a^{(1)},T_b^{(3)}\bigr\}_{(\eta)}
\nonumber\\
&&\phantom{\bigl\{T_a^{(0)},T_b^{(2)}\bigr\}_{(q,p)}}
+\bigl\{T_a^{(2)},T_b^{(2)}\bigr\}_{(\eta)}
+\bigl\{T_a^{(3)},T_b^{(1)}\bigr\}_{(\eta)}=0
\label{A10}\\
&&\phantom{\bigl\{T_a^{(0)},T_b^{(2)}\bigr\}_{(q,p)}+}
\vdots
\nonumber
\end{eqnarray}

\noindent 
Here the notations $\{,\}_{(q,p)}$ and $\{,\}_{(\eta)}$, represent the
parts of the Poisson bracket $\{,\}$ corresponding to the variables
$(q,p)$ and $(\eta)$, respectively. The
equations above are used iteratively to obtain the
corrections $T^{(n)}$ ($n\geq1$). Equation~(\ref{A8}) gives
$T^{(1)}$. Using this result together with the Eq.~(\ref{A9}), 
one calculates $T^{(2)}$, and so on. Since $T^{(1)}$ is linear 
in $\eta$ one can write it as

\begin{equation}
T_a^{(1)}=X_{ab}(q,p)\,\eta^b,
\label{A11}
\end{equation}

\noindent where $X_{ab}$ are some new quantities.
Substituting this expression into (\ref{A8}) and using
(\ref{A1}) and (\ref{A2}), we obtain

\begin{equation}
\Delta_{ab}+X_{ac}\,\omega^{cd}\,X_{bd}=0.
\label{A12}
\end{equation}

\noindent 
We notice that this equation does not define $X_{ab}$ in a unique way,
because it also contains the still unknown elements $\omega^{ab}$. 
What is usually done is to choose $\omega^{ab}$ in such a way that the new
variables are unconstrained. Consequently, the
consistency of the method requires an introduction of other new
variables in order to transform these constraints into the
first class ones. This may lead to an endless process. It is
important to emphasize that $\omega^{ab}$ can be fixed anyway.
However, even if one fixes $\omega^{ab}$, it is still not possible to
obtain a unique solution for $X_{ab}$. Let us check this point. 
$\Delta_{ab}$ and $\omega^{ab}$ are antisymmetric quantities so
expression (\ref{A12}) includes $M(M-1)/2$ independent
equations.  On the other hand, since there is no additional symmetry 
involving $X_{ab}$,  they should represent a set of $M^2$ 
independent quantities.

In the case when $X_{ab}$ does not depend on ($q,p$), it is easily
seen that the expression $T_a+\tilde T_a^{(1)}$ is already strongly 
involutive for any choice we make and we succeed in obtaining 
$\tilde T_a$. If this
is not so, the usual procedure is to introduce $T_a^{(1)}$ into Eq.
(\ref{A9}) in order to calculate $T_a^{(2)}$ and so on. At this 
point one faces a problem that has been the origin of some developments 
of the BFFT method, including the adoption of a non-abelian constraint algebra. This occurs because we do not know {\it a priori} what is the best
choice we can make to go from one step to another. Sometimes it is
possible to figure out a convenient choice for $X_{ab}$ in order to
obtain a first class (abelian) constraint algebra at the first stage
of the process \cite{Banerjee3}. It is opportune to mention
that in ref. \cite{Banerjee4}, the use of a
non-abelian algebra was in fact a way of avoiding to dealing with the higher
orders of the iterative method.

Another point of the usual BFFT formalism is that any dynamic function
$A(q,p)$ (for instance, the Hamiltonian) has also to be properly
modified in order to be strongly involutive with the first class
constraints $\tilde T_a$. Denoting the modified quantity by $\tilde
A(q,p;\eta)$, we then have

\begin{equation}
\bigl\{\tilde T_a,\,\tilde A\bigr\}=0.
\label{A13}
\end{equation}

\noindent
In addition, $\tilde A$ has to satisfy  the boundary condition

\begin{equation}
\tilde A(q,p;0)=A(q,p).
\label{A14}
\end{equation}

\noindent The derivation of $\tilde A$ is similar to what has been 
done in getting $\tilde T_a$. Therefore, we consider an expansion 
of the form

\begin{equation}
\tilde A=\sum_{n=0}^\infty A^{(n)},
\label{A15}
\end{equation}

\noindent 
where $A^{(n)}$ is also a term of order $n$ in $\eta$'s.
Consequently, the compatibility with Eq.~(\ref{A14}) requires that

\begin{equation}
A^{(0)}=A.
\label{A16}
\end{equation}

\noindent 
The combination of Eqs.~(\ref{A6}), (\ref{A7}), (\ref{A13}),
(\ref{A15}), and (\ref{A16}) gives the equations

\begin{eqnarray}
&&\bigl\{T_a,A\bigr\}
+\bigl\{T_a^{(1)},A^{(1)}\bigr\}_{(\eta)}=0
\label{A17}\\
&&\bigl\{T_a,A^{(1)}\bigr\}+\bigl\{T_a^{(1)},A\bigr\}
+\bigl\{T_a^{(1)},A^{(2)}\bigr\}_{(\eta)}
+\bigl\{T_a^{(2)},A^{(1)}\bigr\}_{(\eta)}=0
\label{A18}\\
&&\bigl\{T_a,A^{(2)}\bigr\}
+\bigl\{T_a^{(1)},A^{(1)}\bigr\}_{(q,p)}
+\bigl\{T_a^{(2)},\bigr\}
+\bigl\{T_a^{(1)},A^{(3)}\bigr\}_{(\eta)}
\nonumber\\
&&\phantom{\bigl\{T_a^{(0)},A^{(2)}\bigr\}_{(q,p)}}
+\bigl\{T_a^{(2)},A^{(2)}\bigr\}_{(\eta)}
+\bigl\{T_a^{(3)},A^{(1)}\bigr\}_{(\eta)}=0
\label{A19}\\
&&\phantom{\bigl\{T_a^{(0)},A^{(2)}\bigr\}_{(q,p)}+}
\vdots
\nonumber
\end{eqnarray}

\noindent
which correspond to the coefficients of the powers 0, 1, 2, etc$\dots$  of
the variable $\eta$. It is just a matter of algebraic
work to show that the general expression for $A^{(n)}$ reads as

\begin{equation}
A^{(n+1)}=-{1\over n+1}\,\eta^a\,\omega_{ab}\,X^{bc}\,G_c^{(n)}.
\label{A20}
\end{equation}

\noindent 
where $\omega_{ab}$ and $X^{ab}$ are the inverses of $\omega^{ab}$
and $X_{ab}$, and

\begin{eqnarray}
G_a^{(n)}=\sum_{m=0}^n\bigl\{T_a^{(n-m)},\,A^{(m)}\bigr\}_{(q,p)}
+\sum_{m=0}^{n-2}\bigl\{T_a^{(n-m)},\,A^{(m+2)}\bigr\}_{(\eta)}\nonumber\\
+\bigl\{T_a^{(n+1)},\,A^{(1)}\bigr\}_{(\eta)}.
\label{A21}
\end{eqnarray}

Finally, let us consider the case where the first class
constraints form a non-abelian algebra, i.e. 

\begin{equation}
\bigl\{\tilde T_a,\,\tilde T_b\bigr\}=C_{ab}^c\,\tilde T_c.
\label{A22}
\end{equation}

\noindent
The quantities $C_{ab}^c$ are the structure constants of the
non-abelian algebra. These constraints are considered to satisfy the
same previous conditions given by (\ref{A3}), (\ref{A4}),
(\ref{A6}), and (\ref{A7}). But now, instead of Eqs.
(\ref{A8})-(\ref{A10}), we obtain

\begin{eqnarray}
C_{ab}^c\,T_c&=&\bigl\{T_a,T_b\bigr\}
+\bigl\{T_a^{(1)},T_b^{(1)}\bigr\}_{(\eta)}
\label{A23}\\
C_{ab}^c\,T_c^{(1)}&=&\bigl\{T_a,T_b^{(1)}\bigr\}
+\bigl\{T_a^{(1)},T_b\bigr\}
\nonumber\\
&&+\,\bigl\{T_a^{(1)},T_b^{(2)}\bigr\}_{(\eta)}
+\bigl\{T_a^{(2)},T_b^{(1)}\bigr\}_{(\eta)}
\label{A24}\\
C_{ab}^c\,T_c^{(2)}&=&\bigl\{T_a,T_b^{(2)}\bigr\}
+\bigl\{T_a^{(1)},T_b^{(1)}\bigr\}_{(q,p)}
\nonumber\\
&&+\bigl\{T_a^{(2)},T_b^{(0)}\bigr\}_{(q,p)}
+\bigl\{T_a^{(1)},T_b^{(3)}\bigr\}_{(\eta)}
\nonumber\\
&&+\bigl\{T_a^{(2)},T_b^{(2)}\bigr\}_{(\eta)}
+\bigl\{T_a^{(3)},T_b^{(1)}\bigr\}_{(\eta)+}
\label{A25}\\
&&\vdots
\nonumber
\end{eqnarray}

\noindent 
The use of these equations is the same as before, i.e., they shall
work iteratively. Equation (\ref{A23}) gives $T^{(1)}$.  With this
result and Eq. (\ref{A24}) one calculates $T^{(2)}$, and so on. To
calculate the first correction, we assume it is given by the same
general expression (\ref{A11}). Introducing it into (\ref{A23}), we
now get

\begin{equation}
C_{ab}^c\,T_c=\Delta_{ab}+X_{ac}\,\omega^{cd}\,X_{bd}.
\label{A26}
\end{equation}

\noindent 
Of course, the same difficulties concerning the
solutions of Eq.~(\ref{A12}) also apply here, with the additional
problem of choosing the appropriate structure constants $C_{ab}^c$.
To obtain the embedding Hamiltonian $\tilde H(q,p,\eta)$ one cannot
use the simplified version discussed for the abelian case (embodied
into Eq.(\ref{A22}) ) because the algebra is not strong involutive
anymore. Thus we start from the fact that the new Hamiltonian $\tilde
H$ and the new constraints $\tilde T_a$ satisfy the relation

\begin{equation}
\bigl\{\tilde T_a,\,\tilde H\bigr\}=B_a^b\,\tilde T_b,
\label{A27}
\end{equation}

\noindent where the coefficients $B_a^b$ are the 
structure constant of the non-abelian algebra. The involutive 
Hamiltonian is considered to 
satisfy the same conditions\break(\ref{A14})-(\ref{A16}). We then obtain
that the general correction $H^{(n)}$ is given by a relation similar
to (\ref{A20}), but now the quantities $G_a^{(n)}$ are given by

\begin{eqnarray}
G_a^{(n)}&=&\sum_{m=0}^n\bigl\{T_a^{(n-m)},\,H^{(m)}\bigr\}_{(q,p)}
+\sum_{m=0}^{n-2}\bigl\{T_a^{(n-m)},\,A^{(m+2)}\bigr\}_{(\eta)}
\nonumber\\
&&+\,\,\bigl\{T_a^{(n+1)},\,A^{(1)}\bigr\}_{(\eta)}
-B_a^b\,T_c^{(n)}.
\label{A28}
\end{eqnarray}

\section{Appendix B: the vacuum functional of the model and
the Lagrangian that\\ corresponds to the non-abelian first class
Hamiltonian}

\renewcommand{\theequation}{B\arabic{equation}}
\setcounter{equation}{0}

In this section, we intend to find the Lagrangian that leads
to this new theory. A consistent way of doing this is by
means of the  path integral formalism, where the Faddeev
procedure \cite{FS} has to be used. Let us identify the new 
variables $\eta^\alpha$ as a canonically conjugate pair $ (\phi, 
\pi_\phi)$ in the Hamiltonian formalism,

\begin{eqnarray}
\label{B1}
\eta^1 \rightarrow 2 \phi \,, \nonumber \\
\eta^2 \rightarrow \pi_\phi \,,
\end{eqnarray}

\noindent satisfying (\ref{3.6}). 
Then, the general expression for the vacuum functional reads

\begin{equation}
\label{B19}
Z = N \int [d\mu] \exp \{ i \int dt [ \dot{x}_ip_i
+ \dot{\phi}\pi_\phi - \tilde{H} ] \},
\end{equation}

\noindent with the measure $[d\mu]$ given by

\begin{eqnarray}
\label{B20}
[d\mu] = [dx_i] [dp_i] [d\phi] [d\pi_\phi]
| det\{,\} | \nonumber \\ \delta(x_ix_i-R^2+2R\phi)
\delta(x_ip_i- R\pi_\phi +2\phi\pi_\phi)\prod_\alpha
\delta(\tilde{\Lambda}_\alpha),
\end{eqnarray}

\noindent where $\tilde{\Lambda}_\alpha$ are the gauge fixing 
conditions corresponding to the first class constraints 
$\tilde{T}_\alpha$ and the term $| det\{,\} |$ represents
the determinant of all constraints of the theory, including
the gauge-fixing ones. The quantity N that appears in 
(\ref{B19}) is an usual normalization factor.  Substituting the 
Hamiltonian (\ref{3.15}) into (\ref{B19}), the vacuum functional 
reads

\begin{eqnarray}
\label{B21}
Z = N \int [dx_i] [dp_i] [d\phi] [d\pi_\phi] 
| det\{,\} | \, \delta( x_ix_i-R^2+2R\phi )\nonumber\\
\delta(x_ip_i-R\pi_\phi+2\phi \pi_\phi)
\prod_\alpha
\delta(\tilde{\Lambda}_\alpha)\exp \{ i \int dt 
[ \dot{x}_ip_i + \dot{\phi}\pi_\phi
\nonumber\\
-{1\over 2} p_ip_i(1-{2\phi\over R})
+ x_ip_i {\pi_\phi\over R}(1-{2\phi\over R})
-{1\over 2} x_i x_i {\pi_\phi^2\over R^2}(1-{2\phi\over R})]\}.
\end{eqnarray}

\vskip 1 cm

\noindent Using the delta functions $\delta(x_ix_i-R^2+2R\phi)$ and $\delta(x_ip_i-R\pi_\phi+2\phi \pi_\phi)$, and
exponentiating the last one
with Fourier variable $\xi$, we obtain

\begin{eqnarray}
\label{B22}
Z = N \int [dx_i] [dp_i] [d\phi] [d\pi_\phi] [d\xi]
| det\{,\} | \, \delta( x_ix_i-R^2+2R\phi ) \prod_\alpha
\delta(\tilde{\Lambda}_\alpha) \nonumber \\ \exp \{ i \int dt 
[ \dot{x}_ip_i 
-{1\over 2}p_i^2 (1-{2\phi\over R})+\xi x_i p_i \nonumber \\
+{1\over 2} (1-{2\phi\over R})^2 \pi_\phi^2
-(\dot{\phi}-\xi R(1-{2\phi\over R}))\pi_\phi)]\}.
\end{eqnarray}

\noindent Integrating over $\pi_\phi$, we arrive at

\begin{eqnarray}
\label{B23}
Z = N \int [dx_i] [dp_i] [d\phi] [d\xi]
| det\{,\} | \, \delta( x_ix_i-R^2+2R\phi ) \prod_\alpha
\delta(\tilde{\Lambda}_\alpha) \nonumber \\ 
{1\over 1-{2\phi\over R}}\exp \{ i \int dt 
[ -{1\over 2}(1-{2\phi\over R})p_i^2 + (\dot{x}_i+x_i\xi)p_i \nonumber\\
-{1\over 2} {\dot{\phi}\dot{\phi}\over (1-{2\phi\over R})^2}
+ {R\dot{\phi}\over (1-{2\phi\over R})}\xi
- {1\over 2}R^2\xi^2
 ]\}.
\end{eqnarray}

\vskip 1 cm

\noindent Performing the integration over $p_i$, we obtain

\begin{eqnarray}
\label{B24}
Z = N \int [dx_i][d\phi] [d\xi]
| det\{,\} | \, \delta( x_ix_i-R^2+2R\phi ) \prod_\alpha
\delta(\tilde{\Lambda}_\alpha) \nonumber \\ 
{1\over 1-{2\phi\over R}}
\sqrt{1\over 1-{2\phi\over R}}
\exp \{ i \int dt 
[ 
 {1\over 2}{\dot{x}_i \dot{x}_i\over (1-{2\phi\over R})}
-{1\over 2} {\dot{\phi}\dot{\phi}\over (1-{2\phi\over R})^2} \nonumber \\
+ (1-{2\phi\over R})(x_i \dot{x}_i + R \dot{\phi})\xi
]\}.
\end{eqnarray}

\noindent Finally, the integration over $\xi$ leads to

\begin{eqnarray}
\label{B25}
Z = N \int [dx_i][d\phi] 
|det\{,\}| \, \delta(x_ix_i-R^2+2R\phi) \delta(x_i\dot{x}_i + R\dot{\phi}) \prod_\alpha
\delta(\tilde{\Lambda}_\alpha) \nonumber \\ 
\sqrt{1\over 1-{2\phi\over R}}
\exp  \{ i \int dt [
{1\over 2}{\dot{x}_i \dot{x}_i\over 1-{2\phi\over R}}
-{1\over 2} {\dot{\phi}\dot{\phi}\over (1-{2\phi\over R})^2}] \},
\end{eqnarray}

\noindent where the new $\delta$ function that appear into the expression (\ref{B25}) was obtained
after  integration over $\xi$.
We notice that it does not represent any new restriction over the coordinates of the theory and leads to a consistency condition of constraint $T_1$. From the vacuum functional (\ref{B25}), we identify the extended Lagrangian 

\begin{equation}
\label{B26}
L ={1\over 2}{\dot{x}_i\dot{x}_i\over (1-{2\phi\over R}) }
-{1\over 2} {\dot{\phi}\dot{\phi}\over (1-{2\phi\over R})^2}\,\,.
\end{equation}

\noindent Putting the extended variables, in the phase space,
$\phi$ and $\pi_\phi$ equal to zero, we obtain the original Lagrangian. This result indicates the consistency of the theory.

\end{document}